\documentclass[preprint,floatfix,aps,prd,showpacs,nofootinbib,amsmath,amssymb,amsfonts,superscriptaddress]{revtex4-1}
\usepackage{bm}
\usepackage{graphicx,color}
\usepackage{physics}
\usepackage{dsfont}
\usepackage[normalem]{ulem}
\usepackage{hyperref,url}
\usepackage{mathrsfs}
\usepackage{calrsfs}

\usepackage{subfig}

\usepackage{varioref}
\usepackage[active]{srcltx}

\expandafter\ifx\csname package@font\endcsname\relax\else
\expandafter\expandafter
\expandafter\usepackage
\expandafter\expandafter
\expandafter{\csname package@font\endcsname}%
\fi
\hypersetup{
	colorlinks=true,       
	linkcolor=blue,          
	citecolor=blue,        
}

\usepackage[section]{placeins}

\usepackage{fancybox}
\labelformat{figure}{Fig.~#1}

\begin{document}
\title{Log to log-log crossover of entanglement in $(1+1)-$ dimensional massive scalar field}

\author{Parul Jain} 
\email{paruljain@iitb.ac.in}
\affiliation{Department of Physics, Indian Institute of Technology Bombay, Mumbai 400076, India}

\author{S. Mahesh Chandran\footnote{Corresponding Author}} 
\email{maheshchandran@iitb.ac.in}
\affiliation{Department of Physics, Indian Institute of Technology Bombay, Mumbai 400076, India}
\author{S. Shankaranarayanan}
\email{shanki@phy.iitb.ac.in}
\affiliation{Department of Physics, Indian Institute of Technology Bombay, Mumbai 400076, India}
\begin{abstract}
We study three different measures of quantum correlations --- entanglement spectrum, entanglement entropy, and logarithmic negativity--- for (1+1)-dimensional massive scalar field in flat spacetime. The entanglement spectrum for the discretized scalar field in the ground state indicates a cross-over in the zero-mode regime, which is further substantiated by an analytical treatment of both entanglement entropy and logarithmic negativity. The exact nature of this cross-over depends on the boundary conditions used --- the leading order term switches from a $\log$ to $\log-\log$ behavior for the Periodic and Neumann boundary conditions. In contrast, for Dirichlet, it is the parameters within the leading $\log-\log$ term that are switched. We show that this cross-over manifests as a change in the behavior of the leading order divergent term for entanglement entropy and logarithmic negativity close to the zero-mode limit. We thus show that the two regimes have fundamentally different information content. Furthermore, an analysis of the ground state fidelity shows us that the region between critical point $\Lambda=0$ and the crossover point is dominated by zero-mode effects, featuring an explicit dependence on the IR cutoff of the system. For the reduced state of a single oscillator, we show that this cross-over occurs in the region $Nam_f\sim \mathscr{O}(1)$.
\end{abstract}

\maketitle
\section{Introduction}

Quantum correlations play an important role when describing quantum physics as they help us extract relevant information about a system via measurements.
Quantum correlations are major tools for quantum information, quantum communication, high precision measurements, etc. While there are many ways to measure quantum correlations, one of the most extensively used measures, particularly in field theory, is entanglement entropy \cite{2008Amico.etalRev.Mod.Phys.,2010-Eisert.etal-Rev.Mod.Phys.}.

There is a natural way to partition quantum fields by splitting the degrees of freedom into separate spatial regions. The entanglement entropy can be used to quantify the quantum correlations between the two spatial regions.
To be more precise, the theory can be written on a lattice, with the Hilbert space being a product of Hilbert spaces for each lattice point, i. e. 
${\cal H} = \otimes_i {\cal H}_i$. Let ${\cal H}_A$ be the product of Hilbert spaces at lattice sites within the spatial region A, and ${\cal H}_B$ be the product over the remaining lattice sites so that 
${\cal H} = {\cal H}_A \otimes {\cal H}_B$. Hence, the entanglement entropy associated with a region in some state of the theory can now be determined using quantum mechanical definitions~\cite{1993-Srednicki-Phys.Rev.Lett.,Das2010}.

There are many different approaches to evaluate entanglement entropy for quantum fields: First, as mentioned above, entanglement entropy can be obtained considering the density matrix of a ground state and then tracing out the degrees of freedom confined inside a region. It was shown that in such a case, the entanglement entropy is proportional to the area of the sphere\cite{1993-Srednicki-Phys.Rev.Lett.}. Second, which is what we use in this work, is to exploit the covariance matrix to calculate the entanglement entropy~\cite{2014-Mallayya.etal-Phys.Rev.D}. Third, entanglement entropy is also calculated using the Green's function on a plane and imposing the desired boundary conditions on the finite interval~\cite{2005Casini.HuertaJ.Stat.Mech.,2009Casini.HuertaJ.Phys.A}. This method uses the symmetries of the Helmholtz equation by studying the singular points in the presence of the boundary conditions. With the help of this analysis, one can get $\log Z$ in terms of the solution of a non-linear differential equation of the second-order and the Painlev\'{e} V type. Using this solution, the partition function can be extracted in terms of the correlators of the exponential operators of the Sine-Gordon model. Finally, the replica trick is useful to obtain entanglement entropy for conformal field theories~\cite{2009-Calabrese.Cardy-JournalofPhysicsAMathematicalandTheoretical}. 

At the leading order, all these approaches lead to divergent entanglement
entropy. The divergent term is regulated either using an ultraviolet cutoff or an infrared cutoff.  Depending on the number of space-time dimensions and boundary conditions, the subleading terms to entanglement entropy can also be divergent or non-divergent. In the case of conformal field theories in 
$(1+1)-$dimensions, the subleading term is a non-universal constant~\cite{2005Casini.HuertaJ.Stat.Mech.,2009Casini.HuertaJ.Phys.A,2014-Mallayya.etal-Phys.Rev.D,2009-Calabrese.Cardy-JournalofPhysicsAMathematicalandTheoretical,2016Bianchini.CastroAlvaredoNucl.Phys.B}. 

However, in the case of $(1+1)-$dimensional field theories, the nature of the divergent term can be either $\log$ or $\log-\log$~\cite{2005Casini.HuertaJ.Stat.Mech.,2009Casini.HuertaJ.Phys.A,2014-Mallayya.etal-Phys.Rev.D,2009-Calabrese.Cardy-JournalofPhysicsAMathematicalandTheoretical,2016Bianchini.CastroAlvaredoNucl.Phys.B}. While it is known that the 
presence of the large number of near zero-modes contribute to the 
divergence of the entanglement entropy~\cite{2005Casini.HuertaJ.Stat.Mech.,2009Casini.HuertaJ.Phys.A,2014-Mallayya.etal-Phys.Rev.D,2020Chandran.ShankaranarayananPhys.Rev.D}, it is still unknown why 
certain approaches lead to $\log$ divergence, and other approaches 
lead to $\log-\log$ divergence. 
To elaborate, the authors in Ref.~\cite{2014-Mallayya.etal-Phys.Rev.D}  came across this $\log-\log$ term analytically as a diverging contribution towards the entanglement entropy in the case of periodic boundary conditions. However, the earlier works did not establish an exact relationship between this term and the physical parameters describing the system. On the other hand, for Neumann and Dirichlet boundary conditions in Ref.~\cite{2020Chandran.ShankaranarayananPhys.Rev.D}, the authors could numerically extract a leading log divergent term in place of a log-log term for the entanglement entropy. Further in both Refs.~\cite{2014-Mallayya.etal-Phys.Rev.D,2020Chandran.ShankaranarayananPhys.Rev.D} there were no signs of a crossover with respect to the leading divergent term in the zero-mode regime. In this work, we provide an explicit connection between the results in Refs.~\cite{2014-Mallayya.etal-Phys.Rev.D,2020Chandran.ShankaranarayananPhys.Rev.D} as we analytically obtain a crossover in the leading divergent term of entanglement entropy around $Nam_f\sim \mathscr{O}(1)$, from log to $\log-\log$. This crossover is unique owing to the fact that i) it has not been observed or discussed before in literature, and ii) it is separate from the quantum criticality at $\Lambda=0$ (as discussed in detail below in Sec. ~\ref{overlap}).

We show this crossover by considering two other measures of quantum entanglement --- entanglement negativity and entanglement spectrum. 
Entanglement negativity is the preferred measure to capture entanglement for mixed systems. This is because in dealing with mixed states, entanglement entropy fails to separate the quantum and classical correlations. Negativity involves the sum of the absolute
value of the negative eigenvalues of $\rho_A$ and additionally, one can also calculate the logarithmic negativity, which gives an upper bound in the case of distillable entanglement.
Negativity can be calculated in field theories using a modified replica trick which involves partial transpose of the reduced density matrix. Like 
entanglement entropy, negativity also contains divergent terms~\cite{2013Calabrese.etalJ.Stat.Mech.,2016Bianchini.CastroAlvaredoNucl.Phys.B}.

Entanglement spectrum (ES), corresponding to the eigenvalues of the reduced density matrix, can be used to extract detailed information about the system. For instance, in the case of fractional quantum Hall states, the low-lying levels of entanglement spectrum capture information about the edge modes that help identify topological order, as well as the CFT associated with it~\cite{2008Li.HaldanePhys.Rev.Lett.,2014Chandran.etalPhys.Rev.Lett.}. The difference between the lowest two levels in the spectrum, known as the ``entanglement gap", further contains signatures of symmetry-breaking and quantum phase transitions in many-body systems~\cite{2008Li.HaldanePhys.Rev.Lett.,2012-Ghosh.Shankaranarayanan-PRD,2017-Kumar.Shankaranarayanan-SRep}.
Closing of this gap is found to be associated with quantum criticality~\cite{2020Wald.etalPhys.Rev.Research}.

We explicitly show that the entanglement spectrum of 
$(1+1)-$dimensional massive scalar field in flat space-time hints at a crossover for a certain combination of the parameters. We also establish the relationship between the $\log$ to $\log-\log$ crossover and the presence of zero-modes. To further understand this, we put forth an analytical treatment of the crossover that primarily involves studying the leading order divergent term in the zero-mode limit for entanglement entropy and logarithmic negativity for maximally entangled pure states. The exact nature of this crossover further depends on the boundary conditions used --- the leading order term switches from an overall $\log$ to $\log-\log$ behavior for the Periodic and Neumann boundary conditions, whereas for Dirichlet, the parameters within the leading $\log-\log$ term are switched. We further show that this crossover is a quintessential property of the ground state wave-function by studying the overlap function, a measure that is often used in literature to capture signatures of quantum phase transitions in many-body systems~\cite{1967AndersonPhys.Rev.Lett.,2006Zanardi.PaunkoviifmmodeacutecelsecfiPhys.Rev.E,2008Zhou.BarjaktarevicJournalofPhysicsAMathematicalandTheoretical,2010VieiraJournalofPhysicsConferenceSeries,2017-Kumar.Shankaranarayanan-SRep}.

The paper is organized as follows: In Section \ref{model} we introduce the model and the quantifying tools employed. In Section \ref{espec}, we numerically obtain the entanglement spectrum of the reduced density matrix, which hints at a crossover in the zero-mode regime. To investigate the crossover, we develop the covariance matrix approach to finding entanglement entropy in Section \ref{covar}. In Section \ref{entent}, we use this approach to analyze the leading order divergent contribution in the zero-mode limit for entanglement entropy in the large $N$ limit. Since for maximally entangled pure states, the entanglement entropy is equal to logarithmic negativity, we use this equality to extend the large $N$ entanglement entropy analysis of zero-mode divergence towards logarithmic negativity in Section \ref{negat}. In Section \ref{overlap}, we capture the crossover using the overlap of the ground state wave-function. In Section \ref{conc}, we conclude by discussing the physical interpretations of this crossover, as well as directions of future research. Throughout this work, we use natural units $\hbar=c=k_B=1$.

\section{Model and Quantifying Tools}\label{model}
The Hamiltonian of a massive scalar field in $(1+1)-$dimensions is given by:
\begin{eqnarray}
	H=\frac{1}{2}\int dx \left[\pi^2+(\nabla\varphi)^2+m_f^2\varphi^2\right]
\end{eqnarray}
where $m_f$ is the mass of the scalar field. To evaluate the real-space entanglement entropy of the scalar field, we discretize the above Hamiltonian into a chain of harmonic oscillators by imposing a UV cut-off $a$ as well as an IR cutoff $L=(N+1)a$. On employing a mid-point discretization procedure, the resultant Hamiltonian takes the following form~\cite{Das2010}: 
\begin{equation}
	\label{Ham}
	H=\frac{1}{2a}\sum_j\left[\pi_j^2+\Lambda\varphi_j^2+(\varphi_j-\varphi_{j+1})^2\right]=\frac{1}{a}\tilde{H},
\end{equation}
where 
\begin{equation}
\Lambda=a^2m_f^2 \, .
\label{def:Lambda}
\end{equation}
From its definition, it is clear that $\Lambda$ is invariant under the scaling $(\eta)$ transformations:
\begin{equation}\label{scaling}
	a\to \eta a;\quad m_f\to\eta^{-1}m_f
\end{equation} 
We can then factorize the original Hamiltonian into a scale-dependent part ($1/a$) and a scale-independent part ($\tilde{H}=aH$). This scale-independent Hamiltonian $\tilde{H}$ corresponds to a harmonic chain with nearest neighbor coupling, and can be written as follows:
\begin{equation}
	\label{eq:HC-Hamil}
	\tilde{H}=\frac{1}{2}\left[\sum_i \pi_i^2+\sum_{ij}\varphi_iK_{ij}\varphi_j\right] \, .
\end{equation}
$K_{ij}$ is the coupling matrix that contains relevant information about quantum correlations. The exact form of $K$ depends on the boundary conditions used. The analytical and numerical results for Periodic boundary conditions (PBC) have been extensively discussed in the literature, particularly in the context of zero-modes~\cite{2014-Mallayya.etal-Phys.Rev.D}. This work will focus primarily on the Dirichlet (DBC) and Neumann boundary conditions (NBC), which are much less explored.

To quantify these correlations, we must first calculate the eigenspectrum of the reduced density matrix (RDM) of the subsystem. Given a particular form of coupling matrix $K$, this can be obtained through a well-known procedure \cite{1986-Bombelli.etal-Phys.Rev.D,1993-Srednicki-Phys.Rev.Lett.}. The eigenvalues can then be used to visualize the entanglement spectrum~\cite{2008Li.HaldanePhys.Rev.Lett.,2017-Kumar.Shankaranarayanan-SRep} of the reduced subsystem. Subsequently, they can also be used to calculate the entanglement entropy of the subsystem, which is a popular measure for such correlations. To calculate the entanglement entropy as given by the von Neumann formula, we first consider a bipartite Hilbert space such that we have $\mathcal{H}=\mathcal{H}_A\otimes\mathcal{H}_B$ wherein $\mathcal{H}_A$ pertains to the subsystem $A$ and $\mathcal{H}_B$ pertains to the subsystem $B$. We next consider a pure state $|\Psi\rangle$ such that the density matrix is $\rho = |\Psi\rangle\langle \Psi|$ and the reduced density matrix is $\rho_A=\mathrm{Tr}_B\rho$ to finally get the entanglement entropy as~\cite{2009-Calabrese.Cardy-JournalofPhysicsAMathematicalandTheoretical}
\begin{equation}\label{EE}
	S_A=-\mathrm{Tr}\rho_A\ln\rho_A\, . 
\end{equation}

For the above model, it has been shown that the ground state entanglement entropy corresponding to $H$ and $\tilde{H}$ are related as~\cite{2020Chandran.ShankaranarayananPhys.Rev.D}:
\begin{eqnarray}
	S=\tilde{S}(\Lambda)
\end{eqnarray}
Hence it is sufficient to work with the rescaled Hamiltonian $\tilde{H}$. 

In general, computing the entanglement entropy from RDM even for a single oscillator reduced state requires numerical implementation. Alternately, we can also arrive at the entanglement entropy by considering the covariance matrix of the system\cite{2010-Eisert.etal-Rev.Mod.Phys.}. In this approach, it is possible to obtain analytical expressions for the entropy for the reduced state of a single oscillator\cite{2014-Mallayya.etal-Phys.Rev.D}. Therefore, in this work we rely on the covariance matrix approach to obtain the leading order term of entanglement entropy. Here, the quantum vacuum state is a Gaussian state~\cite{2008Amico.etalRev.Mod.Phys.,2010-Eisert.etal-Rev.Mod.Phys.,2014-Mallayya.etal-Phys.Rev.D}. A Gaussian state is defined as:
\begin{equation}
W(x,p)\propto e^{-\frac{1}{2}(R-\langle R\rangle)^T\sigma^{-1}(R-\langle R\rangle)}
\end{equation}
where $\sigma$ is the covariance matrix given by
\begin{equation}
\sigma_{kl}=\frac{1}{2}\langle R_k R_l + R_l R_k \rangle-\langle R_k\rangle \langle R_l \rangle
\end{equation}
and $R=(X_{1}, X_{2},...,X_{N},P_{1},P_{2},...,P_{N})^{\dagger}$. In the nomenclature of distribution function, an $N$-mode Gaussian state is characterized by the $2N$-dimensional covariance matrix $\sigma$ and the $2 N$-dimensional first moments. The covariance matrix for an N-mode Gaussian state is of the form:
\[ \sigma = \left[ \begin{array}{cc}
\sigma_{XX} & \sigma_{XP} \\
\sigma_{PX} & \sigma_{PP} \end{array} \right].\]
The partial trace on a Gaussian state is also a Gaussian state with reduced number of modes. The covariance matrix of this subsystem can be constructed by picking the variances of those modes in the total covariance matrix that belong to the reduced subsystem. The entanglement entropy depends only on the covariance matrix.

While entanglement entropy serves as a good measure to capture entanglement for pure states, it fails when it comes to mixed states, in which case it is unable to separate the quantum and classical contributions. For a mixed state, we hence rely on a more general measure to capture such correlations, such as entanglement negativity. Entanglement negativity is given as \cite{2002Vidal.WernerPhys.Rev.A,2014Rangamani.RotaJHEP}:
\begin{equation}
	\mathcal{N}(\rho) = \frac{\|\rho^{\Gamma}\| - 1}{2}, 
\end{equation}
where $\rho^{\Gamma}$ is the partial transpose of the density matrix $\rho$ and
$\|\rho^{\Gamma}\|$ is the trace norm and it is the sum of the absolute values of the eigenvalues of $\|\rho^{\Gamma}\|$ meaning $\|\rho^{\Gamma}\| = \mathrm{Tr}|\rho^{\Gamma}|$. Next, we can say
\begin{equation}
	\mathrm{Tr}(\rho^{\Gamma}) = \sum\limits_i \lambda_i^{(+)} + \sum\limits_j \lambda_j^{(-)} \equiv 1 = \mathrm{Tr}(\rho),
\end{equation}
Using the above equation, we can then define negativity as
\begin{equation}
	\mathcal{N}(\rho) = \frac{1}{2}\Bigg( \sum\limits_i |\lambda_i^{(+)}| + \sum\limits_j |\lambda_j^{(-)}| - 1 \Bigg) = 
	\sum\limits_j |\lambda_j^{(-)}|,
\end{equation} 
which indeed shows that negativity is the sum of the absolute values of the negative eigenvalues of $\rho$. 

We can further define what is called as the {\it{logarithmic negativity}} as 
\begin{equation}
	\mathcal{E_N}(\rho) = \log\|\rho^{\Gamma}\|, 
\end{equation}
which serves as an upper bound for the distillable entanglement. Further,  we have
\begin{equation}\label{47}
	\mathcal{E_N}(\rho) = S(\rho), 
\end{equation}
for a maximally entangled pure state.

\section{Entanglement Spectrum}\label{espec}

To capture purely quantum correlations in the field between two sub-regions, it is sufficient to obtain the reduced density matrix (RDM) by tracing out the degrees of freedom corresponding to a sub-region. Reduced density matrix contains complete information about quantum entanglement; however, entanglement entropy being scalar may not provide complete information~\cite{2008Li.HaldanePhys.Rev.Lett.,2017-Kumar.Shankaranarayanan-SRep}.
The entanglement spectrum of the reduced system is defined as:
\begin{equation}
	h_E=-\log{\rho_{red}}
\end{equation}
Here, we consider a chain of $2N$ coupled harmonic oscillators that simulate the properties of the scalar field, and trace out all oscillators but one---the $N^{th}$ oscillator in the chain. We do this to minimize the edge effects in the system, as well as for direct comparison with the analytical results obtained in Sections \ref{covar} and \ref{entent}.
\begin{figure*}[!ht]
	\begin{center}
		\subfloat[\label{es1a}][]{%
			\includegraphics[width=0.4\textwidth]{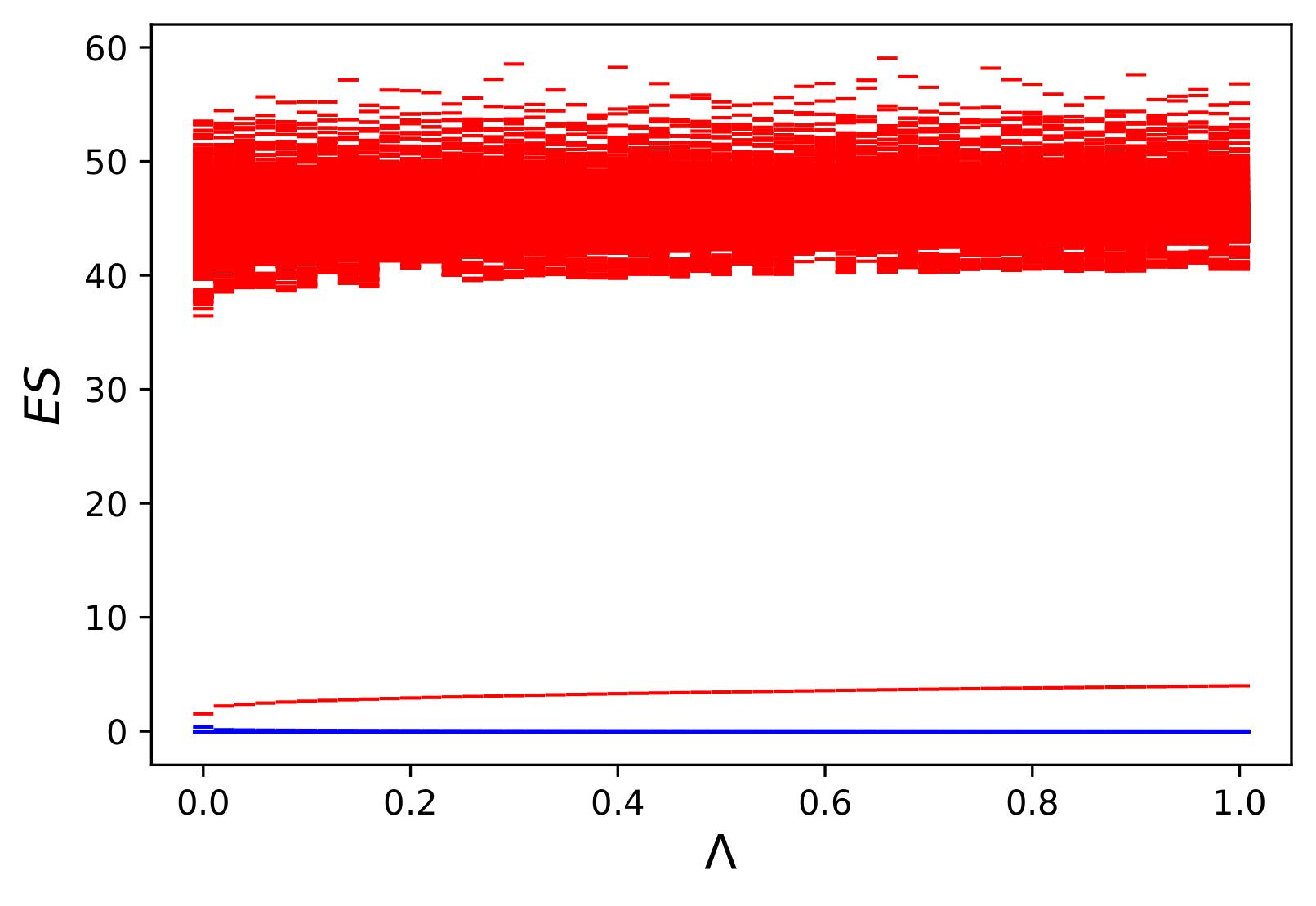}
		}
		\subfloat[\label{es1b}][]{%
			\includegraphics[width=0.4\textwidth]{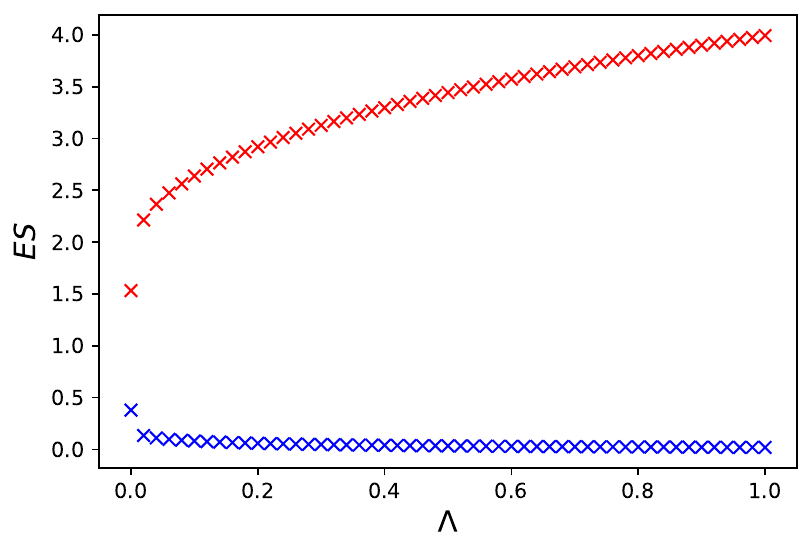}
		}
		
		\caption{(a) Entanglement spectrum (ES) and (b) Entanglement gap for $n=0$ (blue) and $n=1$ (red) eigenvalues of the reduced density matrix for DBC. Here, $N=1000$.}
		\label{es1}
	\end{center}
\end{figure*}

In harmonic chains, the largest eigenvalues of reduced density matrix correspond to $n=0$, and $n=1$ \cite{1993-Srednicki-Phys.Rev.Lett.}. We keep track of the effective gap between these two levels by looking at the largest values corresponding to $n=0$ and the smallest values corresponding to $n=1$. We call this the ``entanglement gap". Depending on the boundary conditions, we see that both the spectrum and gap have a characteristic behavior on varying the rescaled mass $\Lambda$~(cf. Eq \ref{def:Lambda}) of the scalar field. 

\begin{figure*}[!ht]
	\begin{center}
		\subfloat[\label{es2a}][]{%
			\includegraphics[width=0.4\textwidth]{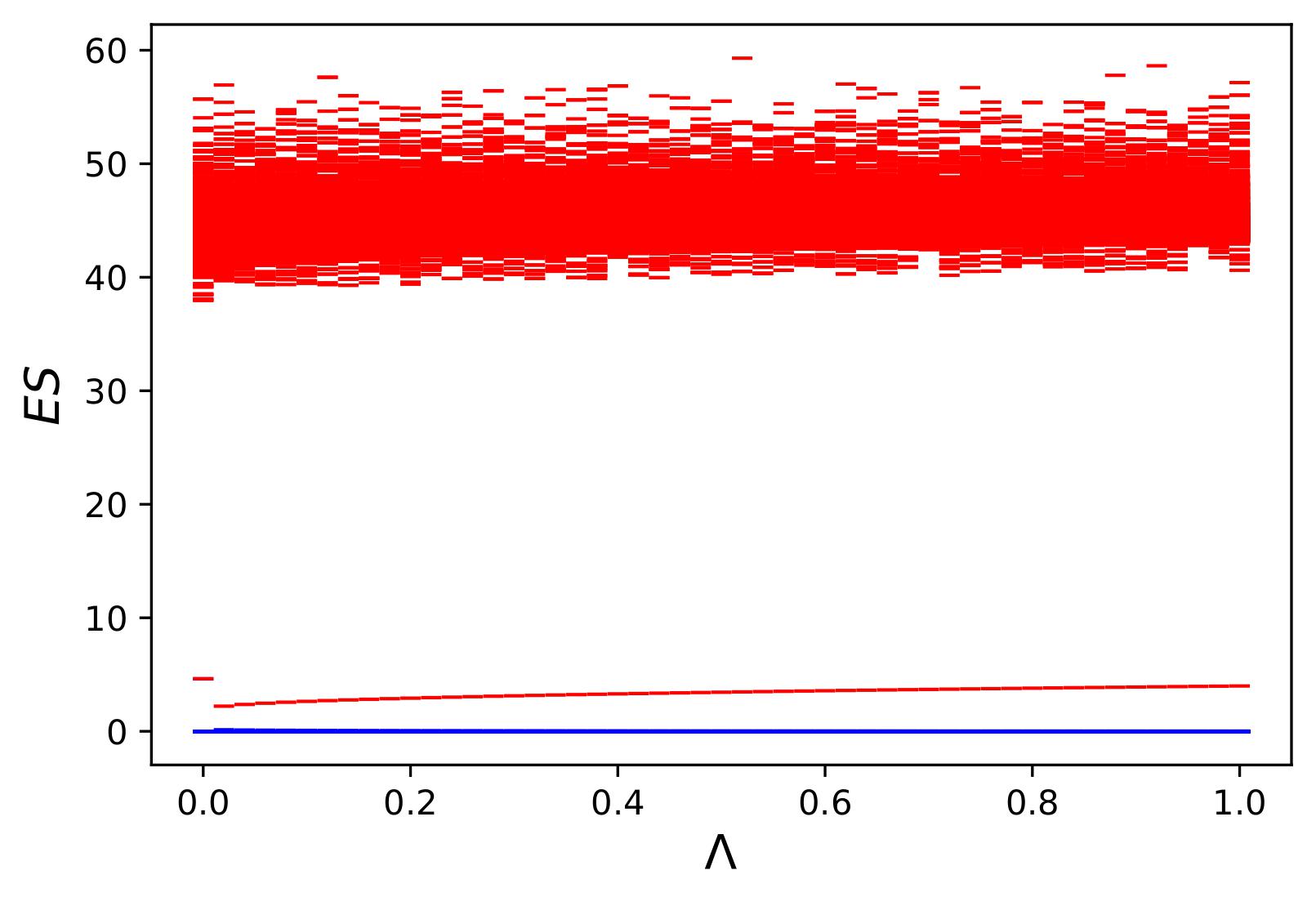}
		}
		\subfloat[\label{es2b}][]{%
			\includegraphics[width=0.4\textwidth]{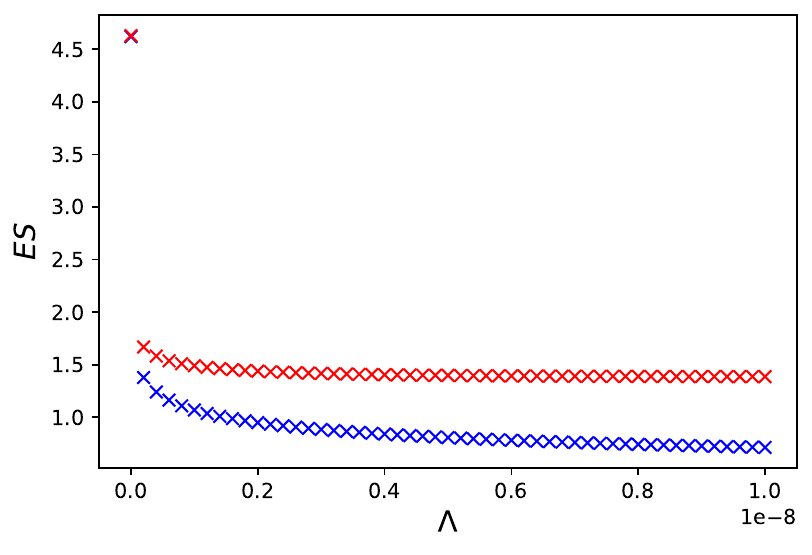}
		}
		
		\caption{Entanglement spectrum (ES) and (b) Entanglement gap for $n=0$ (blue) and $n=1$ (red) eigenvalues of the reduced density matrix for NBC. Here, $N=1000$.}
		\label{es2}
	\end{center}
\end{figure*}

As seen in both \ref{es1} and \ref{es2}, the nearby levels seem to draw closer as $\Lambda\to0$, which is also the limit associated with zero-mode divergence of entanglement entropy. While the levels seemingly converge in this limit for NBC, there remains a distinct gap for DBC. However, we know that while NBC always has a zero-mode for any value of $N$, DBC can only generate zero-modes in the limit $N\to\infty$ \cite{2020Chandran.ShankaranarayananPhys.Rev.D}. We, therefore, expect this convergence for DBC only in the thermodynamic limit. This also establishes a strong connection between degeneracy in entanglement spectra and zero-mode divergence of entanglement entropy. 

From this analysis, we, therefore, observe that the entanglement gap seemingly closes near the limit $\Lambda\to0$ and widens as $\Lambda$ increases. This hints at a possible crossover between two regimes with fundamentally different information content for some combination of the parameters that describe the system, namely $N$, $a$, and $m_f$. It has been previously noted in the literature that we obtain a critical point as $N\to\infty$, $m_f\to0$ and $a\to0$, corresponding to the conformal limit of a $(1+1)-$dimensional scalar fields \cite{2006CALABRESE.CARDYInternationalJournalofQuantumInformation}. However, the model we have taken here is finite with a well-defined UV cut-off and a non-zero mass. In the rest of this work, we will try to understand what causes this crossover. We will also try to obtain a fundamental understanding of what these two regimes indicate and how they are connected to zero-modes~\cite{2020Chandran.ShankaranarayananPhys.Rev.D}.

\section{Covariance Matrix Approach to Entanglement Entropy}\label{covar}
To investigate the crossover hinted at in Section \ref{espec}, we look at other measures that capture quantum correlations in the system. In this section, we rely on the covariance matrix approach to obtain exact analytical expressions of entanglement entropy for the case of a single oscillator subsystem. The rescaled Hamiltonian $\tilde{H}$ defined in \eqref{Ham} corresponds to a chain of harmonic oscillators with nearest-neighbor coupling. For $N$ such oscillators, the covariance matrix is a $2N\cross2N$ matrix given by\cite{2010-Eisert.etal-Rev.Mod.Phys.}:
\begin{equation}
	\sigma=\frac{1}{2}\begin{bmatrix}
		K^{-1/2}&0\\0&K^{1/2}
	\end{bmatrix}=\frac{1}{2}\begin{bmatrix}
	A&0\\0&B
\end{bmatrix},
\end{equation}
where $K$ is the coupling matrix whose elements are fixed depending on the boundary conditions as well as the parameters $\Lambda$ and $N$. 
\subsection{Dirichlet Boundary Condition}
In this subsection, we impose the condition $\varphi_0=\varphi_{N+1}=0$. The coupling matrix $K_{ij}$ becomes a symmetric Toeplitz matrix with the following non-zero elements:
	\begin{equation}
		K_{jj}=\Lambda+2;\quad K_{j,j+1}=K_{j+1,j}=-1
	\end{equation}
The normal modes are calculated to be~\cite{2008-Willms-SIAMJournalonMatrixAnalysisandApplications}:
\begin{eqnarray}
		\tilde{\omega}_k^2=\Lambda+4\cos^2{\frac{k\pi}{2(N+1)}}\quad k=1,..N
\end{eqnarray}
	We immediately see that the system does not develop any zero-modes even when $\Lambda=0$ as long as $N$ is finite. In the thermodynamic limit ($N \to \infty$), the Dirichlet chain develops exactly one zero-mode ($\tilde{\omega}_N$) and a large number of near-zero-modes. The normalized eigenvectors are given by:
	\begin{equation}
		v_j^{(m)}=\sqrt{\frac{2}{N+1}}\sin{\left(\frac{jm\pi}{N+1}\right)}=M_{jm},
	\end{equation}
	where $M_{jm}$ is the diagonalizing matrix such that $MKM=diag\{\tilde{\omega}_j\}$. The elements of the covariance matrix are therefore:
	\begin{align}
		A_{lm}&=\frac{1}{N+1}\sum_{j=1}^N\frac{1}{\tilde{\omega}_j}\sin{\left(\frac{lj\pi}{N+1}\right)}\sin{\left(\frac{jm\pi}{N+1}\right)}\nonumber\\
		B_{lm}&=\frac{1}{N+1}\sum_{j=1}^N\tilde{\omega}_j\sin{\left(\frac{lj\pi}{N+1}\right)}\sin{\left(\frac{jm\pi}{N+1}\right)}
	\end{align}
	For a single-oscillator reduced system, the reduced covariance matrix can be obtained by picking appropriate elements from the total covariance matrix\cite{2019Giulio.etalJournalofStatisticalMechanicsTheoryandExperiment}. For simplicity, let us consider the $N^{th}$ oscillator in a system of $2N$ oscillators. The reduced covariance matrix is of the form:
	\begin{equation}
		\sigma_{red}=\frac{1}{2}\begin{bmatrix}
			A_{NN}&0\\0&B_{NN}
		\end{bmatrix}
	\end{equation}
	The determinant of the reduced covariance matrix is given by:
	\begin{equation}
		\det{\sigma_{red}}=\frac{1}{\left(2N+1\right)^2}\sum_{i=1}^{2N}\frac{\sin^2{\left(\frac{iN\pi}{2N+1}\right)}}{\sqrt{\Lambda+4\cos^2{\left(\frac{i\pi}{4N+2}\right)}}}\sum_{j=1}^{2N}\sin^2{\left(\frac{jN\pi}{2N+1}\right)}\sqrt{\Lambda+4\cos^2{\left(\frac{j\pi}{4N+2}\right)}}
	\end{equation}
	For large enough $N$, we see that:
	\begin{equation}
		\sin^2{\left(\frac{iN\pi}{2N+1}\right)}\approx \sin^2{\left(\frac{i\pi}{2}\right)}=\begin{cases}0\quad\text{i is even}\\1\quad\text{i is  odd}\end{cases}
	\end{equation}
	As a result, the determinant can be simplified as follows: 
	\begin{equation}\label{dbcdet}
	\det{\sigma_{red}}\approx \frac{1}{\left(2N+1\right)^2}\sum_{k=1}^{N}\frac{1}{\sqrt{\Lambda+4\cos^2{\left(\frac{(2k-1)\pi}{4N+2}\right)}}}\sum_{l=1}^{N}\sqrt{\Lambda+4\cos^2{\left(\frac{(2l-1)\pi}{4N+2}\right)}}
	\end{equation}
	
\subsection{Neumann Boundary Condition}
We impose the condition $\partial_x \varphi=0$ at the two ends of the chain by setting $\varphi_0=\varphi_1$ and $\varphi_{N+1}=\varphi_N$. The resultant coupling matrix is, therefore, a perturbed symmetric Toeplitz matrix whose non-zero elements are given below:
\begin{equation}
		K_{jj\neq1,N}=\Lambda+2;\quad K_{11}=K_{NN}=\Lambda+1;\quad K_{j,j+1}=K_{j+1,j}=-1
\end{equation}
The normal modes (eigenvalues of $K$) are found to be~\cite{2008-Willms-SIAMJournalonMatrixAnalysisandApplications}:
\begin{eqnarray}
		\tilde{\omega}_k^2=\Lambda+4\cos^2{\frac{k\pi}{2N}};\quad k=1,..,N
\end{eqnarray}
We see that the system develops exactly one zero-mode ($\tilde{\omega}_N$) when $\Lambda=0$, even for a finite $N$. The normalized eigenvectors are given by:
\begin{equation}
	v_j^{(m)}=\begin{cases}\sqrt{\frac{2}{N}}\sin{\left(\frac{(2j-1)m\pi}{2N}\right)}\quad m=1,..,N-1\\\frac{(-1)^{j-1}}{\sqrt{N}}\qquad\qquad\qquad\, m=N\end{cases}
\end{equation}
The elements of the covariance matrix are therefore:
\begin{align}
	A_{lm}&=\frac{1}{N\sqrt{\Lambda}}+\frac{2}{N}\sum_{j=1}^{N-1}\frac{1}{\tilde{\omega}_j}\sin{\left[\left(l-\frac{1}{2}\right)\frac{j\pi}{2N}\right]}\sin{\left[\left(m-\frac{1}{2}\right)\frac{j\pi}{2N}\right]}\nonumber\\
	B_{lm}&=\frac{\sqrt{\Lambda}}{N}+\frac{2}{N}\sum_{j=1}^{N-1}\tilde{\omega}_j\sin{\left[\left(l-\frac{1}{2}\right)\frac{j\pi}{2N}\right]}\sin{\left[\left(m-\frac{1}{2}\right)\frac{j\pi}{2N}\right]}
\end{align}
Let us again consider the reduced state of the $N^{th}$ oscillator in a system of $2N$ oscillators. The reduced covariance matrix is of the form:
\begin{equation}
	\sigma_{red}=\frac{1}{2}\begin{bmatrix}
		A_{NN}&0\\0&B_{NN}
	\end{bmatrix}
\end{equation}
For large enough $N$, similar to what was done for the Dirichlet case, the determinant of the reduced covariance matrix can be simplified as follows:
\begin{equation}\label{nbcdet}
	\det{\sigma_{red}}\approx \frac{1}{4N^2}\left[\frac{1}{2\sqrt{\Lambda}}+\sum_{k=1}^{N}\frac{1}{\sqrt{\Lambda+4\cos^2{\left(\frac{(2k-1)\pi}{4N}\right)}}}\right]\left[\frac{\sqrt{\Lambda}}{2}+\sum_{l=1}^{N}\sqrt{\Lambda+4\cos^2{\left(\frac{(2l-1)\pi}{4N}\right)}}\right]
\end{equation}
From this, we can calculate the entanglement entropy for the single-oscillator subsystem as follows: 
\begin{equation}
	S=\left(\alpha+\frac{1}{2}\right)\log{\left(\alpha+\frac{1}{2}\right)}-\left(\alpha-\frac{1}{2}\right)\log{\left(\alpha-\frac{1}{2}\right)},
\end{equation}
where $\alpha=\sqrt{\det{\sigma_{red}}}$. If the determinant (and hence $\alpha$) is very large, we may simplify the expression as follows:
\begin{equation}
	S\approx\log{\alpha}=\frac{1}{2}\log{\left(\det{\sigma_{red}}\right)}
\end{equation}

\section{Entanglement Entropy: zero-mode divergence and normal mode spacing}\label{entent}
In this section, we analyze the leading order terms of entanglement entropy and probe for a crossover in the zero-mode regime. While the approach used in Appendix \ref{appa} sufficiently captures this crossover, we take a slightly different route so as to obtain a better physical insight. Let us consider low-lying normal modes in a system of $2N$ oscillators. For DBC, when $N$ is sufficiently large, we see that:
\begin{equation}
	\tilde{\omega}_{2N-1}^2=\Lambda+4\cos^2{\left(\frac{(2N-1)\pi}{4N+2}\right)}=\Lambda+4\sin^2{\left(\frac{\pi}{2N+1}\right)}\approx\Lambda+\frac{\pi^2}{N^2}
	\end{equation}
Let us now consider the relative spacing of the lowest two normal modes with respect to the rescaled mass gap $\Lambda$, defined by $\zeta$:
\begin{equation}
	\zeta_{DBC}=\frac{\tilde{\omega}_{2N-1}^2-\tilde{\omega}_{2N}^2}{\Lambda}\approx \frac{3\pi^2}{4N^2\Lambda}
\end{equation}
The quantity defined above can also be represented differently depending on the parameters we wish to tune:
\begin{equation}
	\zeta_{DBC}\approx\frac{3\pi^2}{4N^2\Lambda}=\frac{3}{4}\left(\frac{\pi}{Nam_f}\right)^2=\frac{3}{4}\left(\frac{\pi}{Lm_f}\right)^2
\end{equation}
Similarly, for very large $N$ in the case of NBC, we get:
\begin{equation}
	\zeta_{NBC}\approx \frac{\pi^2}{4N^2\Lambda}
\end{equation}
We can see from here that the relative spacing for Dirichlet is three times that of Neumann. Ideally, we would like to consider $a \ll 1$ and $N \gg 1$. However, the relative speeds of taking these limits lead to varying behavior in $\zeta$. The following limits of $\zeta$ are relevant:
\begin{itemize}
	\item $\zeta\ll1$ : Small relative level spacing. Corresponds to the case when $a\to0$ or $m_f\to0$ is slower than $N\to\infty$. The former is also equivalent to the limit $L\to\infty$.
	\item $\zeta\gg 1$ : Large relative level spacing. Corresponds to the case when $a\to0$ or $m_f\to0$ is faster than $N\to\infty$. The former is also equivalent to the limit $L\to0$.
\end{itemize}
We now show that the above two limits lead to vastly different behavior in entanglement entropy of the system.
\subsection{Small relative level spacing $\zeta\ll 1$}
\begin{figure*}[!ht]
	\begin{center}
		\subfloat[\label{dir2a}][]{%
			\includegraphics[width=0.4\textwidth]{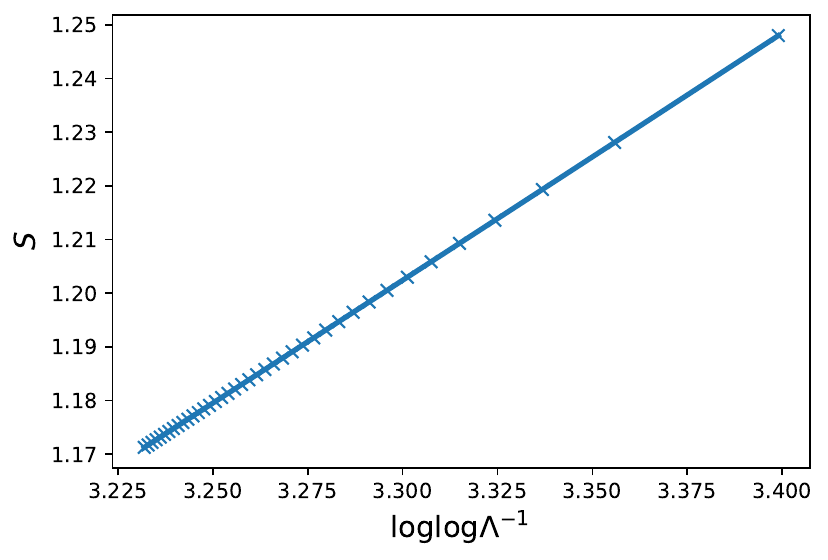}
		}
		\subfloat[\label{neu2a}][]{%
			\includegraphics[width=0.4\textwidth]{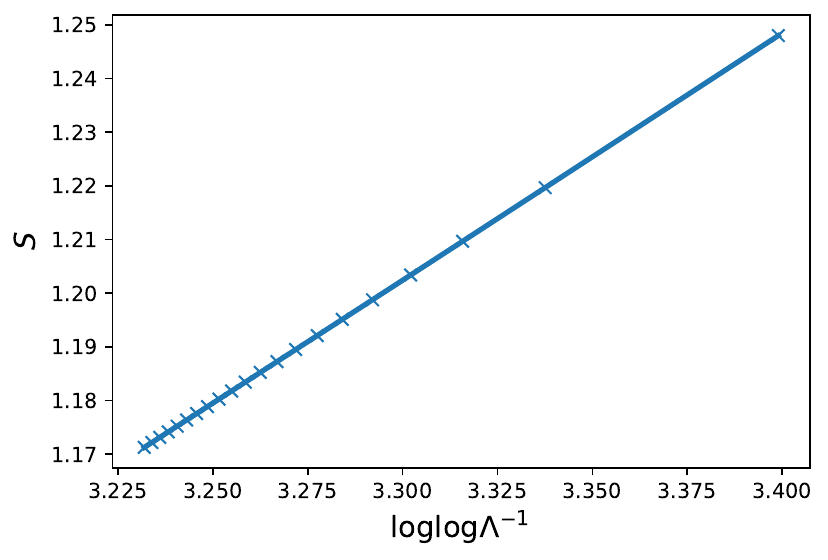}
		}
		
		\caption{Entanglement scaling for a) DBC and b) NBC when $N=10^8$ and $\Lambda\in(10^{-13},10^{-11})$ corresponding to $\zeta\ll1$.}
		\label{dirneu}
	\end{center}
\end{figure*}

In this limit, we can replace the summation in \eqref{dbcdet} and \eqref{nbcdet} with  an integral, since the spacing is almost continuous. For DBC, we can introduce $\theta=(2k-1)\pi/(4N+2)$, as a result of which:
\begin{align}
	\det{\sigma_{red}}&\approx \frac{1}{\pi^2}\int_0^{\frac{\pi}{2}}\frac{d\theta}{\sqrt{\Lambda+4\cos^2{\theta}}}\int_0^{\frac{\pi}{2}}\sqrt{\Lambda+4\cos^2{\theta}}d\theta\nonumber\\
	&=\frac{1}{\pi^2}\int_0^{\frac{\pi}{2}}\frac{d\theta}{\sqrt{1-k^2\sin^2{\theta}}}\int_0^{\frac{\pi}{2}}\sqrt{1-k^2\sin^2{\theta}}d\theta \nonumber\\
	&=\frac{1}{\pi^2}K(k)E(k),
\end{align}
where $K$ and $E$ are complete elliptic integrals with modulus $k^2=4/(\Lambda+4)$~\cite{1971-Byrd.Friedman-HandbookEllipticIntegrals}. It should be noted that in this limit, the determinant becomes independent of the number of oscillators $N$. On expanding the above expression upto the leading order in $\Lambda$, we get:
\begin{equation}
	\det{\sigma_{red}}\approx \frac{1}{2\pi^2}\log{\frac{64}{\Lambda}}+\mathscr{O}(\Lambda\log\Lambda)
\end{equation}
The determinant therefore diverges as $\Lambda\to0$. The leading order contribution to entanglement entropy is therefore:
\begin{equation}
	\lim_{\zeta\to0}S_{DBC}\sim \frac{1}{2}\log{\log{\left(\frac{64}{\Lambda}\right)}}
\end{equation} 
Entanglement entropy diverges due to the presence of zero-mode (since we are taking $N\to\infty$), but the divergence is slow. The $\log-\log$ divergence, as we will see, is exclusive to the case $\zeta\ll1$, which can be attained by taking $N\to\infty$ faster than $\Lambda\to0$. On performing a similar analysis for Neumann, we see that:
\begin{equation}
	\lim_{\zeta\to0}S_{NBC}\sim \frac{1}{2}\log{\log{\left(\frac{64}{\Lambda}\right)}}\sim \lim_{\zeta\to0}S_{DBC}
\end{equation} 
As can be seen in \ref{dirneu}, we therefore obtain the same behavior of entanglement entropy for both Neumann and Dirichlet, in the limit $\zeta\ll 1$.

\subsection{Large relative level spacing $\zeta\gg 1$}
We know that the limit $\zeta\gg 1$ corresponds to:
\begin{equation}
	\zeta_{DBC} \approx \frac{3\pi^2}{4N^2\Lambda}\gg 1\implies \Lambda\ll \frac{3\pi^2}{4N^2}
\end{equation}
From \eqref{dbcdet}, we see that $\Lambda$ is negligible compared to $\cos^2\left(\pi/(4N+2) \right)$ in the determinant for DBC. In the limit $\zeta\to\infty$, we may therefore ignore $\Lambda$ and leads to:
\begin{equation}
	\det{\sigma_{red}}\approx \frac{1}{\left(2N+1\right)^2}\sum_{k=1}^{N}\sec{\left(\frac{(2k-1)\pi}{4N+2}\right)}\sum_{l=1}^{N}\cos{\left(\frac{(2l-1)\pi}{4N+2}\right)}
\end{equation}
It is difficult to obtain a closed form expression for the secant summation. However, keeping with the limit $\zeta\gg 1$, we may take the limit $N\to\infty$ slower than $\Lambda\to0$, and hence replace the summation with an integral.
\begin{align}
	\det{\sigma_{red}}&\approx \frac{1}{\pi^2}\int_0^{\frac{\pi}{2}-\frac{\pi}{2N}}\sec{\theta}d\theta\int_0^{\frac{\pi}{2}-\frac{\pi}{2N}}\cos{\theta}d\theta\nonumber\\
	&=\frac{1}{\pi^2}\log{\left(\csc{\frac{\pi}{2N}}+\cot{\frac{\pi}{2N}}\right)}\nonumber\\
	&\approx \frac{1}{\pi^2}\log{\left(\frac{4N}{\pi}\right)}+\mathscr{O}(N^{-2})
\end{align}
The leading order contribution to entanglement entropy for the single-oscillator subsystem is:
\begin{equation}
	\lim_{\zeta\to\infty}S_{DBC}\approx\frac{1}{2}\log{\log{\left(\frac{4N}{\pi}\right)}}
\end{equation}

\begin{figure}[!hbt]
	\centering
	\includegraphics[scale=0.5]{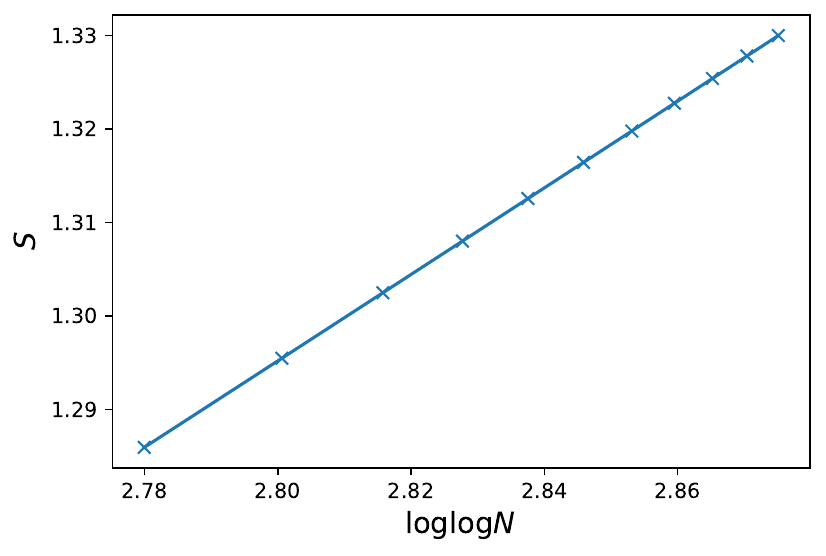}
	\caption{Entanglement scaling for DBC when $N\in(10^{7},5\cross10^{7})$ and $\Lambda=0$, corresponding to $\zeta\gg1$.}
	\label{dir2}
\end{figure}
For DBC, we see that even when $\Lambda=0$, the entropy does not diverge unless $N\to\infty$. However, this divergence is very slow, but unlike the case where $\zeta\ll 1$, it depends on $N$ instead of $\Lambda$. This implies that on taking $N\to\infty$ and $\Lambda\to0$, the divergence is effectively determined by the slower limit. 

\begin{figure*}[!ht]
	\begin{center}
		\subfloat[\label{neu1a}][$\Lambda=10^{-16}$ and $N\in(500,1000)$]{%
			\includegraphics[width=0.4\textwidth]{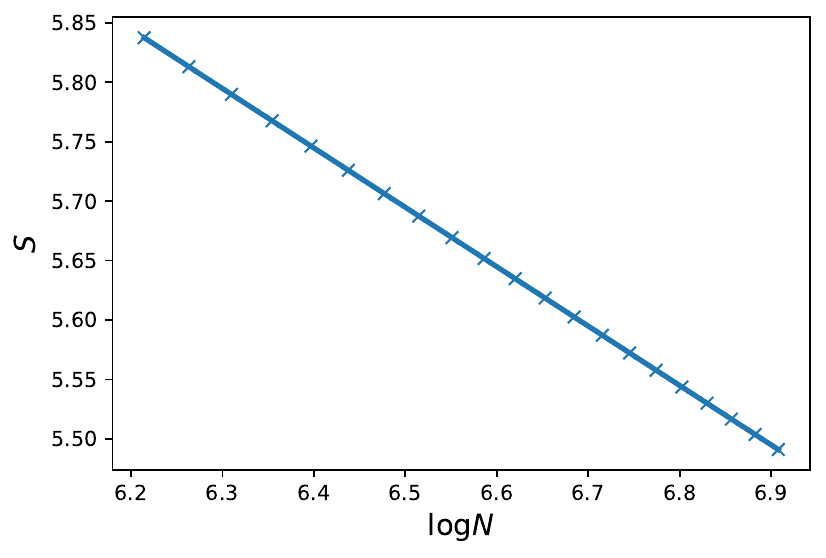}
			
		}
		\subfloat[\label{neu1b}][$\Lambda\in(10^{-18},10^{-16})$ and $N=10^4$.]{%
			\includegraphics[width=0.4\textwidth]{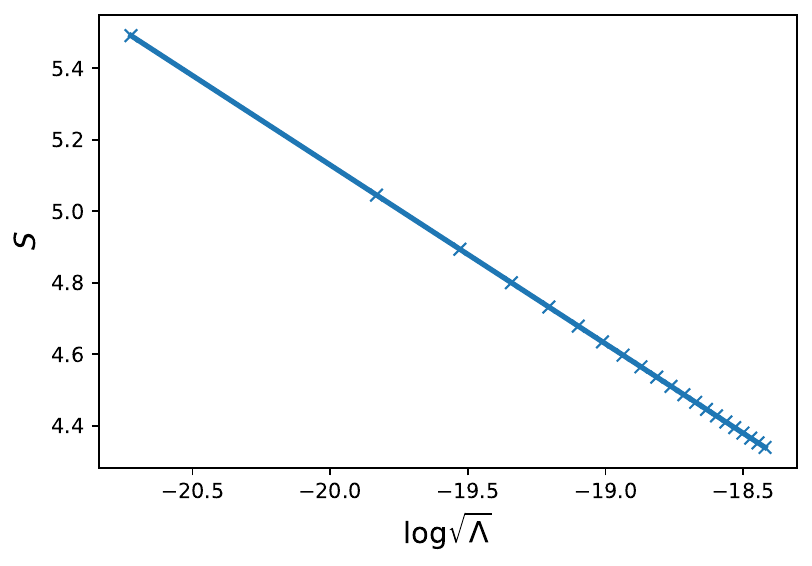}
		}
		
		\caption{Entanglement scaling for NBC with respect to a) $\log{N}$ and b) $\log{am_f}$ when $\zeta\gg1$.}
		\label{neu2}
	\end{center}
\end{figure*}

Let us now perform a similar analysis on NBC by assuming $\Lambda$ is negligible compared to the cosine term inside the square root in \eqref{nbcdet}:
\begin{equation}
	\det{\sigma_{red}}\approx \frac{1}{16N^2}\left[\frac{1}{\sqrt{\Lambda}}+\sum_{k=1}^{N}\sec{\left(\frac{(2k-1)\pi}{4N}\right)}\right]\left[\sqrt{\Lambda}+4\sum_{l=1}^{N}\cos{\left(\frac{(2l-1)\pi}{4N}\right)}\right]
\end{equation}
Here again, it is difficult to obtain a closed form expression for the secant summation. Hence, like in DBC, we assume that $N$ is large enough for the summation to be replaced by an integral such that $\zeta\to\infty$:
\begin{align}
	\det{\sigma_{red}}&\approx \frac{1}{4N^2}\left[\frac{1}{2\sqrt{\Lambda}}+\frac{2N}{\pi}\int_0^{\frac{\pi}{2}-\frac{\pi}{4N}}\sec{\theta}d\theta\right]\left[\frac{\sqrt{\Lambda}}{2}+\frac{2N}{\pi}\int_0^{\frac{\pi}{2}-\frac{\pi}{4N}}\cos{\theta}d\theta\right]\nonumber\\
	&=\left[\frac{1}{4N\sqrt{\Lambda}}+\frac{1}{2\pi}\log{\left(\csc{\frac{\pi}{4N}}+\cot{\frac{\pi}{4N}}\right)}\right]\left[\frac{\sqrt{\Lambda}}{4N}+\frac{2}{\pi}\cos{\frac{\pi}{4N}}\right]\nonumber\\
	&\approx \frac{1}{2\pi N\sqrt{\Lambda}}+\frac{1}{\pi^2}\log{\left(\frac{8N}{\pi}\right)}+\mathscr{O}(N^{-1}\log{N})
\end{align}
From the above expression, it is clear that the determinant diverges as $\Lambda\to0$ even for a finite $N$, unlike what is observed for DBC. The leading order contribution to entanglement entropy for the single-oscillator subsystem is:
\begin{equation}
	\lim_{\zeta\to\infty}S_{NBC}\approx-\frac{1}{2}\log{\left(N\sqrt{\Lambda}\right)}
\end{equation}
Furthermore, as can be seen in \ref{dir2} and \ref{neu2}, we can conclude that in the limit $\zeta\gg1$ the leading order term is sensitive to boundary conditions.

\begin{table}[!htb]
	\centering
	\resizebox{0.8\textwidth}{!}{%
		\begin{tabular}{@{}|c|c|c|@{}}
			\toprule
			Boundary Condition & Small relative level spacing & Large relative level spacing \\ 
			 & $Nam_f\gg  1$& $Nam_f \ll 1$ \\ 
			\toprule
			Dirichlet & $S\sim\frac{1}{2}\log{\left(-\log{am_f}\right)}$& $S\sim\frac{1}{2}\log{\log{N}} $  \\[6pt] \hline
			Neumann& $S\sim\frac{1}{2}\log{\left(-\log{am_f}\right)}$ & $S\sim -\frac{1}{2}\log{\left(Nam_f\right)}$\\[6pt] \hline
			Periodic & $S\sim\frac{1}{2}\log{\left(-\log{am_f}\right)}$& $S\sim -\frac{1}{2}\log{\left(Nam_f\right)}$  \\[6pt]
			\toprule
		\end{tabular}
	}
	\caption{Summary of leading order contribution to entanglement entropy for a single-oscillator subsystem, in the limits of large $N$ and small $am_f$ ($\log am_f<0$).}
	\label{tab:summary}
\end{table}

\subsection{Effects on scaling symmetry}
For a reduced state of a single oscillator, the subsystem-dependent terms of entanglement entropy are suppressed by the zero-mode divergent terms. This is no longer the case for a larger subsystem size. In the limit $\Lambda\to0$, the subsystem scaling relations for entanglement entropy upto leading order (for both Dirichlet and Neumann) are as follows\cite{2020Chandran.ShankaranarayananPhys.Rev.D}:
\begin{align}
	S \sim \frac{1}{6}\log{\frac{r}{a}}+S^{(1)}
\end{align}
The first term in the above expression is independent of $\Lambda$ and is also invariant under the scaling transformations in \eqref{scaling}. The term $S^{(1)}$, on the other hand, does not depend on subsystem size $r=na$ and is generally treated as sub-leading.  These subleading terms depend on the parameters $\{N,\Lambda\}$ and $\zeta=Nam_f$. As a result, this term is also invariant under the transformations in \eqref{scaling}. This leads us to the conclusion that for very large $N$, entanglement entropy $S$ cannot distinguish between the limits $a\to0$ and $m_f\to0$, which are physically very different. Despite this, there may be a difference in the speed of divergence of entropy in the respective limits, which might help break this degeneracy. To see how this occurs, let us look at the dominant term in entropy:
\begin{itemize}
	\item Continuum limit $a\to0$ : To ensure that both the subsystem size ($r=na$) and full system size ($L=Na$) of the model are non-zero, we must also rapidly take the limit $n,N\to\infty$. As a result, the leading-order divergence will always include the subsystem dependent term $\log{(r/a)}$, irrespective of the behavior of $S^{(1)}$.
	\item Massless limit $m_f\to0$: Here, the subsystem dependent term is finite and does not contribute to entropy divergence. The divergence, therefore, arises from $S^{(1)}$, the nature of which ($\log$ or $\log-\log$) can be inferred from Table \ref{tab:summary}. This also implies that the $S^{(1)}$ is no longer a sub-leading term.
\end{itemize}
From the above analysis, we conclude that the nature of leading order divergence of entropy can, in general, distinguish the limits $a\to0$ (log) and $m_f\to0$ ($\log-\log$).  The exception is when we stick to the limit $Nam_f\ll1$ for the Neumann or Periodic boundary conditions, in which case both the limits give rise to a $\log$ divergence. 


\section{Logarithmic Negativity}\label{negat}
In this section, we evaluate the leading order ground state logarithmic negativity
for a $(1+1)-$dimensional massive scalar field in a flat space-time for periodic, Neumann, and Dirichlet boundary conditions. 
We will consider the system described in \eqref{Ham} in both the finite $N$ and large $N$ limit.

\subsection{Periodic Boundary Conditions : Finite N}
We begin with periodic boundary conditions implying $\varphi_0=\varphi_{N}$. The
dispersion relation in this case is~\cite{2020Chandran.ShankaranarayananPhys.Rev.D}
\begin{equation}
	\tilde{\omega}_k^2=\Lambda+4\sin^2\left(\frac{\pi (j-1)}{N}\right) 
\end{equation}
where $j=1...N$. Now, using the above dispersion relation
we can extract the eigenvalues $(\lambda_N)^2$ from the determinant of the covariance matrix for the single oscillator reduced system, which is given as \cite{2014-Mallayya.etal-Phys.Rev.D} (where we have considered the $N$th oscillator in a chain of 2$N$ oscillators) 
\begin{multline}
	\mathrm{Det(\sigma_{red})} = \frac{1}{16N^2}\left[
	\frac{1}{\sqrt{\Lambda}} + \frac{1}{\sqrt{\Lambda+4}}
	+2\sum_{i=1}^{N-1}\frac{1}{\sqrt{\Lambda+4\sin^2\left(\frac{\pi i}{2N}\right)}}\right]\\\cross
	\left[\sqrt{\Lambda}+\sqrt{\Lambda+4}+
	2\sum_{j=1}^{N-1}\sqrt{\Lambda+4\sin^2\left(\frac{\pi j}{2N}\right)}
	\right]  
\end{multline}
We now recall that logarithmic negativity is given as $\mathcal{E_N}=\sum_j\log|\lambda_j|$ which in the present case will turn out to be $\mathcal{E_N}=\log|\lambda_N|$.
So, we can 
now write $\mathcal{E_N}$ as
\begin{multline}\label{ENNBC}
	\mathcal{E_N} = \frac{1}{2}\log\left|\left[
	\frac{1}{4N}\left[\frac{1}{\sqrt{\Lambda}} + \frac{1}{\sqrt{\Lambda+4}}+2\sum_{i=1}^{N-1}\frac{1}{\sqrt{\Lambda+4\sin^2\left(\frac{\pi i}{2N}\right)}}
	\right]\right]\right|\\
	+\frac{1}{2}\log\left|\left[
	\frac{1}{4N}\left[\sqrt{\Lambda}+\sqrt{\Lambda+4}+2\sum_{j=1}^{N-1}\sqrt{\Lambda+4\sin^2\left(\frac{\pi j}{2N}\right)}\right]\right]\right|  
\end{multline}
Taking $\Lambda=0$ in the above equation leads to the following expression:
\begin{equation}
	\mathcal{E_N} \sim  \frac{1}{2}\log\left|\frac{1}{4N\sqrt{\Lambda}}\right|
\end{equation}
From the above equation, we see that, for $\Lambda=0$ and $N$ finite, 
$\mathcal{E_N}$ has a divergent $\log$ term.

\subsection{Neumann Boundary Conditions: Finite N}
We next consider the Neumann boundary conditions implying $\varphi_0=\varphi_1, \varphi_N=\varphi_{N+1}$ and $\partial_x\varphi=0$. In this case, the dispersion relation is   \cite{2020Chandran.ShankaranarayananPhys.Rev.D}
\begin{equation}
	\tilde{\omega}_k^2=\Lambda+4\cos^2\left(\frac{k \pi}{2N}\right) 
\end{equation}
where $k=1...N$. Now, using this dispersion relation and 
the method similar to the periodic boundary conditions 
the determinant is:
\begin{multline}
	\mathrm{Det(\sigma_{red})} = \frac{1}{4N^2}\left[\sum_{i=1}^{2N-1}\frac{\sin^2\left(\frac{i\pi (2N-1)}{4N}\right)}{\sqrt{\Lambda+4\cos^2\left(\frac{\pi i}{4N}\right)}}+\frac{1}{2\sqrt{\Lambda}}\right]\\\cross\left[\sum_{j=1}^{2N-1}\sin^2\left(\frac{j\pi (2N-1)}{4N}\right)\sqrt{\Lambda+4\cos^2\left(\frac{\pi j}{4N}\right)}+\frac{\sqrt{\Lambda}}{2}\right]  
\end{multline}
Using the above determinant, we can 
now express $\mathcal{E_N}$ as
\begin{multline}
	\mathcal{E_N} = \frac{1}{2}\log\left|\left[\frac{1}{2N}\sum_{i=1}^{2N-1}\frac{\sin^2\left(\frac{i\pi (2N-1)}{4N}\right)}{\sqrt{\Lambda+4\cos^2\left(\frac{\pi i}{4N}\right)}}+\frac{1}{2\sqrt{\Lambda}}\right]\right|\\ +
	\frac{1}{2}\log\left|
	\left[\frac{1}{2N}\sum_{j=1}^{2N-1}\sin^2\left(\frac{j\pi (2N-1)}{4N}\right)\sqrt{\Lambda+4\cos^2\left(\frac{\pi j}{4N}\right)}+\frac{\sqrt{\Lambda}}{2}\right]\right|  
\end{multline}
Taking $\Lambda=0$ in the above equation leads to the following expression:
\begin{equation}
	\mathcal{E_N} \sim  \frac{1}{2}\log\left|\frac{1}{4N\sqrt{\Lambda}}\right|
\end{equation}
Like in the periodic boundary condition, for $\Lambda=0$ and $N$ finite, 
$\mathcal{E_N}$ has a divergent $\log$ term.

\subsection{Dirichlet Boundary Conditions : Finite N}
Finally, we consider  the Dirichlet boundary conditions implying $\varphi_0=\varphi_{N+1}=0$ and wherein the dispersion relation is \cite{2020Chandran.ShankaranarayananPhys.Rev.D}
\begin{equation}
	\tilde{\omega}_k^2=\Lambda+4\cos^2\left(\frac{k \pi}{2(N+1)}\right) 
\end{equation}
where $k=1...N$. Further, the determinant in this case is given as
\begin{small}\begin{equation}
		\mathrm{Det(\sigma_{red})} = \frac{1}{(2N+1)^2}\sum_{i=1}^{2N}\frac{\sin^2\left(\frac{i\pi N}{2N+1}\right)}{\sqrt{\Lambda+4\cos^2\left(\frac{\pi i}{2(2N+1)}\right)}}\sum_{j=1}^{2N}\sin^2\left(\frac{j\pi N}{2N+1}\right)\sqrt{\Lambda+4\cos^2\left(\frac{\pi j}{2(2N+1)}\right)}  
\end{equation}\end{small}
Making use of the above determinant we can 
now write $\mathcal{E_N}$ as
\begin{small}
	\begin{equation}\label{ENDBC}
		\mathcal{E_N} = \frac{1}{2}\log\left|\frac{1}{(2N+1)^2}\sum_{i=1}^{2N}\frac{\sin^2\left(\frac{i\pi N}{2N+1}\right)}{\sqrt{\Lambda+4\cos^2\left(\frac{\pi i}{2(2N+1)}\right)}}\sum_{j=1}^{2N}\sin^2\left(\frac{j\pi N}{2N+1}\right)\sqrt{\Lambda+4\cos^2\left(\frac{\pi j}{2(2N+1)}\right)}  \right| 
	\end{equation}
\end{small}
From the above expression, we see that $\mathcal{E_N}$ is finite for $\Lambda=0$. In periodic and Neumann boundary conditions, 
zero-mode is present for the finite $N$ case. However, in the Dirichlet case,
there is no zero-mode for the finite $N$ case.
\subsection{Large $N$ limit analysis}
We now proceed towards the large $N$ limit of the system and 
perform a similar analysis for negativity as was done for the entanglement entropy to compare the results. We see that in the large $N$ limit, as the covariance matrix remains unchanged after the partial transpose, we can have the same eigenvalues for both the entropy and negativity for the maximally entangled pure state. Further, as we consider only the $N^{th}$ oscillator, we will have $\mathcal{E_N}=\log\lambda_N$. To conclude, in the large $N$ limit, we have,
\begin{equation}
	S=\mathcal{E_N}=\log\lambda_N 
\end{equation}
Further, the above equation validates the equality between entanglement entropy and logarithmic negativity for the maximally entangled pure states \cite{2002Vidal.WernerPhys.Rev.A,2014Rangamani.RotaJHEP}. 

As a result of this equality in the large $N$-limit,  the results  obtained in Sections \ref{espec}, \ref{covar}, and \ref{entent} can be extended to $\mathcal{E_N}$. Appendix \ref{appEN} contains the explicit calculations. 

Before we proceed, we want to compare and contrast the results in this work with the earlier results in the literature. In  Ref.~\cite{2014-Mallayya.etal-Phys.Rev.D}, for the periodic boundary conditions, the authors came across the $\log-\log$ term analytically as a diverging contribution towards the entanglement entropy. However, 
the earlier works did not establish an exact relationship between this term and the physical parameters describing the system. In Ref.~\cite{2020Chandran.ShankaranarayananPhys.Rev.D}, for Neumann and Dirichlet boundary conditions, the authors numerically extracted \emph{only} the leading log-divergent term. Further in both Refs.~\cite{2014-Mallayya.etal-Phys.Rev.D,2020Chandran.ShankaranarayananPhys.Rev.D} there were no signs of a crossover with respect to the leading divergent term in the zero-mode regime. In this work, we have provided an explicit connection between the results in Refs.~\cite{2014-Mallayya.etal-Phys.Rev.D,2020Chandran.ShankaranarayananPhys.Rev.D} as we analytically obtain a crossover in the leading divergent term of entanglement entropy around $Nam_f\sim \mathscr{O}(1)$, from log to log-log. This crossover is unique owing to the fact that i) it has not been observed or discussed before in literature, and ii) As we show in the next section, the crossover is separate from the quantum criticality at $\Lambda=0$.

\section{The ground state overlap function}\label{overlap}

In this section, we look for the crossover beyond entanglement and, especially, in measures that capture the fundamental properties of the ground state wave-function. The overlap function or ground state fidelity captures signatures of phase transitions in various quantum systems~\cite{2006Zanardi.PaunkoviifmmodeacutecelsecfiPhys.Rev.E,2008Zhou.BarjaktarevicJournalofPhysicsAMathematicalandTheoretical,2010VieiraJournalofPhysicsConferenceSeries,2017-Kumar.Shankaranarayanan-SRep}, and therefore can be tested to see if the crossover is an essential feature of the ground state wave-function of a $(1+1)-$D massive scalar field. For an infinitesimal change $\delta \Lambda$ in the value of rescaled mass $\Lambda$, the ground state overlap function can be calculated as follows:
\begin{align}
	F&=\bra{\Psi_0(\Lambda+\delta \Lambda)}\ket{\Psi_0(\Lambda)}\nonumber\\
	&=\frac{\det^{1/4}\Omega(\Lambda) \det^{1/4}\Omega(\Lambda+\delta\Lambda)}{\det^{1/2}\left(\frac{\Omega(\Lambda)+\Omega(\Lambda+\delta\Lambda)}{2}\right)},
\end{align}
where $\Omega=K^{1/2}$. Since the diagonalizing matrix for $\Omega$ is independent of $\Lambda$, it takes the exact same form for both $\Lambda$ and $\Lambda+\delta \Lambda$ cases. As a result, the determinant in the denominator can be simplified as the product of average of corresponding normal modes for both $\Lambda$ and $\Lambda+\delta \Lambda$. Therefore, for a system of $2N$ oscillators, the overlap function further simplifies to:
\begin{equation}
	F=\prod_{k=1}^{2N}F_k=2^N\prod_{k=1}^{2N}\frac{\tilde{\omega}_k^{1/4}(\Lambda) \tilde{\omega}_k^{1/4}(\Lambda+\delta\Lambda)}{\left(\tilde{\omega}_k(\Lambda)+\tilde{\omega}_k(\Lambda+\delta\Lambda)\right)^{1/2}}
\end{equation}

\begin{figure*}[!ht]
	\begin{center}
		\subfloat[\label{over1a}][DBC]{%
			\includegraphics[width=0.4\textwidth]{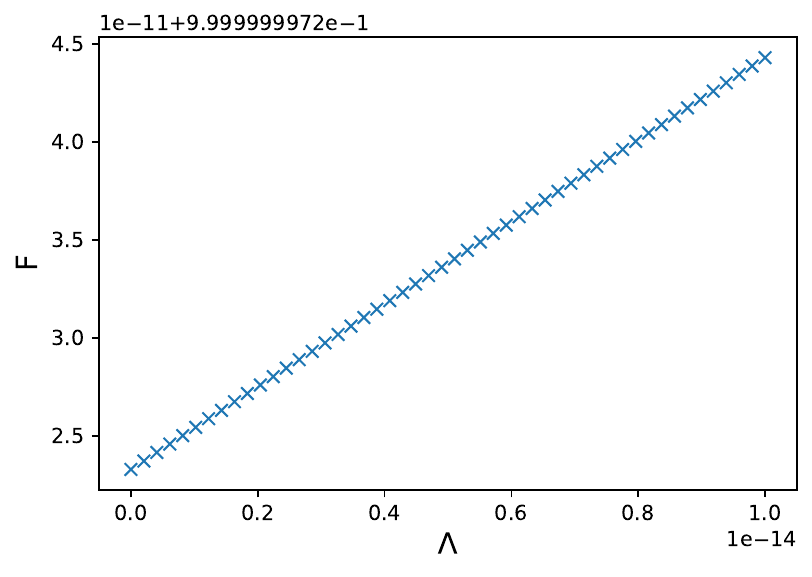}
			
		}
		\subfloat[\label{over1b}][NBC]{%
			\includegraphics[width=0.4\textwidth]{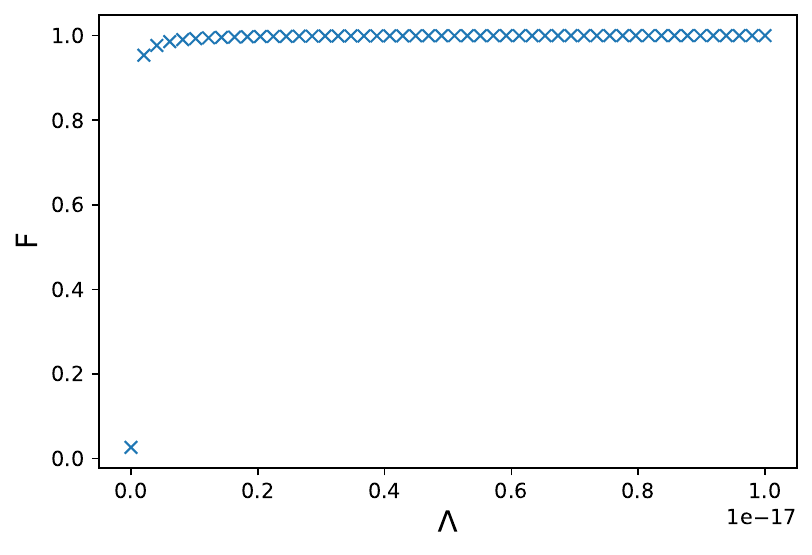}
		}
		
		\caption{Plot of Overlap function as a function of $\Lambda$ for (a) DBC and (b) NBC. We have set $\delta\Lambda=10^{-16}$ and $N=10^6$.}
		\label{over1}
	\end{center}
\end{figure*}

We see that the overlap functions for both DBC and NBC behave quite differently. For DBC, the overlap function remains very close to unity and is expected to approach zero only as $N\to\infty$ when a zero-mode is generated. For NBC, the presence of a zero-mode for a finite $N$ causes the overlap function to fall sharply to zero as $\Lambda\to0$. This merely points out that $\Lambda=0$ leads to orthogonal states in the system for finite (infinite) $N$ for Neumann (Dirichlet). Coupling this with the divergences that develop in the system, such as that of entanglement entropy, $\Lambda=0$ indicates quantum criticality. We would, however, like to know whether this overlap function also captures signatures of a crossover about $Nam_f\sim \mathscr{O}(1)$ as was observed for entanglement entropy in previous sections.

Assuming that the infinitesimal mass shift $\delta\Lambda \ll \tilde{\omega}_{k}^2$, we can expand the individual contributions to overlap function as follows:
\begin{equation}
	F_k\sim 1-\frac{\delta \Lambda^2}{\tilde{\omega}_k^4}
\end{equation}
Let us now consider the smallest $F_k$, which is also the contribution that can indicate when or how rapidly the overlap function falls to zero from unity. We bring back the relative level spacing parameter $Nam_f$ and analyze various limits to obtain the asymptotics as summarized in Table \ref{tab:overlap}. Here, we observe two fundamentally different forms for the overlap function on either side of the crossover, similar to the results we obtained for entanglement entropy. When $Nam_f\ll 1$, the overlap function is determined by the size of the oscillator system $N$. On the other hand, when $Nam_f\gg1$, the overlap function is determined by the rescaled mass $\Lambda$ of the scalar field.
\begin{table}[!htb]
	\centering
	\resizebox{0.7\textwidth}{!}{%
		\begin{tabular}{@{}|c|c|c|@{}}
			\toprule
			Boundary Condition & $Nam_f\ll  1$& $Nam_f \gg 1$  \\ 
			\toprule
			Dirichlet & $F_{2N}\sim1-\frac{N^4\delta \Lambda^2}{4\pi^4}$& $F_{2N}\sim 1-\frac{\delta \Lambda^2}{64\Lambda^2} $  \\[6pt] \hline
			Neumann& $F_{2N-1}\sim1-\frac{N^4\delta \Lambda^2}{4\pi^4}$& $F_{2N-1}\sim 1-\frac{\delta \Lambda^2}{64\Lambda^2} $\\[6pt] 
			\toprule
		\end{tabular}
	}
	\caption{Signature of crossover in the overlap function. It should be noted that the smallest fidelity contribution corresponds to $k=2N$ in DBC and $k=2N-1$ in NBC.}
	\label{tab:overlap}
\end{table}

We must treat the critical point at $\Lambda=0$ and the crossover at $Nam_f\sim \mathscr{O}(1)$ separately. The crossover marks a distinct onset of zero-mode effects --- most notably, the observables develop sensitivity to system size (IR cutoff) $L\sim Na$, which then acts as an additional control parameter for quantum entanglement and quantum fidelity, apart from the mass of the scalar field. This crossover manifests itself a noticeable shift in (i) the leading order divergent term of entanglement entropy and (ii) the individual contributions to the fidelity function for an infinitesimal shift $\delta \Lambda$. The nature of crossover depends on the boundary conditions used. These zero-mode effects amplify on progressing to the critical point $\Lambda=0$, wherein entanglement entropy diverges, entanglement gap closes, and fidelity vanishes, marking complete orthogonality of the states in and around $\Lambda=0$. The critical point and the crossover point converge exactly only at the field theory limit $N\to\infty$ and $\Lambda=0$. However, for a large but finite system size, the crossover point and critical point remain disparate. The region in between exhibits exotic IR-dependent characteristics that are also sensitive to the boundary conditions used. The behaviour exactly at the crossover point or the critical point is beyond the scope of the current work as we need to employ other sophisticated analytical techniques, like DMRG, to obtain more concrete results.


\section{Conclusions and Discussions}\label{conc}

In this work, we have studied an interesting crossover in the zero-mode regime for the ground state of a discretized massive scalar field in $(1+1)$-dimensions. The crossover has signatures across the three measures of quantum correlations. In Section \ref{espec}, we calculated the entanglement spectrum for the ground state reduced density matrix for NBC and DBC systems. For finite $N$ in NBC, we observed a closing of the gap approaching the zero-mode limit $\Lambda\to0$. For DBC, zero-modes only appear when both $\Lambda\to0$ and $N\to\infty$, and hence we did not see a closing of the gap for finite $N$. This result establishes a connection between zero-modes and the closing of the entanglement gap, wherein the latter is generally associated with quantum criticality. This also implies that the entanglement entropy divergence usually arising from zero-modes can be attributed to degeneracy in the lower levels of the entanglement spectrum.

To investigate this crossover further, we looked at the leading order terms of entanglement entropy in the zero-mode regime. By tracing out a single oscillator from the system, we could exert analytic control of the model. In Section \ref{entent}, we introduced a new quantity $\zeta$ that captured the relative spacing between the lowest two normal modes. We showed that in the limits $\zeta\gg1$ and $\zeta\ll1$ near the zero-mode limit, the leading order terms of entanglement entropy reduced to drastically different forms. In the small relative spacing limit $\zeta\ll1$ for NBC and PBC, the zero-mode divergence was slower ($\log-\log$) than the faster log-divergence in the large relative spacing limit $\zeta\gg1$. For DBC, both limits resulted in a slow $\log-\log$ divergence, but the parameters inside the $\log-\log$ term were switched. The exact details of the crossover have been summarized in Table \ref{tab:summary} as well as in Appendix \ref{appa}.

On studying the pure state logarithmic negativity of the system in Section \ref{negat}, we see that when we consider the zero-mode regime in the finite $N$ scenario, we get the leading divergent term scaling as a $\log$ for periodic and Neumann boundary conditions. On the other hand, there is no such divergent term in the Dirichlet boundary condition. As the finite $N$ study reflects no change in the leading divergent term from 
$\log$ to log(log) in the zero-mode limit 
so to extract the crossover signature, we proceed towards the large $N$ limit of the system. To study this case we use the fact that in the large $N$ limit, for pure states, $S=\mathcal{E_N}$ and extend the results as obtained for the entanglement entropy in Sections \ref{espec}, \ref{covar}, and \ref{entent} to logarithmic negativity as well. To conclude, we can say that for pure states, the crossover analysis leads to the same results 
for both entanglement entropy and 
logarithmic negativity.

On tracing out more oscillators, the subsystem-dependent term will no longer be suppressed. In this case, we analyzed the scaling symmetry associated with the transformations $a\to\eta a$ and $m_f\to\eta^{-1}m_f$ that left the entropy invariant. In earlier work, we argued that this symmetry caused the zero-mode divergence arising from $a\to0$ or $m_f\to0$ to be indistinguishable when confining ourselves to the subsystem-dependent term. However, with the inclusion of log/$\log-\log$ terms that depend on full-system parameters, we see that the speed of divergence may be different for the limits $a\to 0$ and $m_f\to0$ in special cases. This suggests that the scaling symmetry mentioned above may be broken for certain limits of the system parameters. This is an interesting problem which we hope to address in later work.

We analytically proved the existence of the crossover that we saw in the entanglement spectrum. While the entanglement gap asymptotically closed in the limit of $\Lambda\to0$, we showed that the parameter that ultimately decided this crossover is $\zeta$, which depends on both system size $N$, and the rescaled scalar field mass $\Lambda=a^2m_f^2$. Suppose we fix $N$ for the system to be very large; we see that the crossover occurs in the region $\Lambda\sim N^{-2}$. On decreasing $\Lambda$ below this threshold, we see that the leading order term picks up from a slower $\log-\log$ behavior to a faster $\log$ behavior for NBC and PBC. We identify this to be the region where the entanglement gap begins to close, wherein the first two levels of entanglement spectra approach degeneracy. We also note that above this threshold, the entropy of all three boundary conditions coincide whereas it is boundary dependent below this threshold. As we look at larger $N$ values, the threshold value of $\Lambda$ becomes smaller. Finally, when we extend the system size to infinity ($N=\infty$), we see that the crossover is possible only at $\Lambda=0$, which corresponds to a critical point the scalar field in $(1+1)$-dimensions\cite{2006CALABRESE.CARDYInternationalJournalofQuantumInformation}.

By studying the overlap function in Section \ref{overlap}, we have shown that the crossover is also a fundamental feature of the ground state wave-function. The crossover point $\Lambda\sim N^{-2}$ marks the onset of zero-mode effects in the system, wherein it develops an explicit dependence on system size (or the IR cutoff), similar to what was observed in entanglement entropy. This is separate from the critical point at $\Lambda=0$, and the region in between these two points is characterized by a sudden development of orthogonality of neighboring quantum states in the parameter space, which would otherwise have been nearly indistinguishable. In the field theory limit, the crossover point and critical point converge, and the overlap function vanishes. We hope to address the IR-dependence and other interesting features exactly at the critical point or the crossover point in later work.

For higher dimensions, we rely on partial wave expansion of the scalar field to reduce the Hamiltonian of the system into an effective $(1+1)$-dimensional form \cite{1993-Srednicki-Phys.Rev.Lett.,2020Chandran.ShankaranarayananPhys.Rev.D}. For $(3+1)$-dimensions, the coupling matrix $K$ corresponding to $l=0$ reduces almost exactly to that of $(1+1)$-dimensions for very large $N$, but deviates drastically for larger values of $l$~\cite{2012-Ghosh.Shankaranarayanan-PRD}. However, the contribution to entanglement entropy is generally dominated by lower values of $l$, particularly the $l=0$ wave that gives rise to a zero-mode in the limit $\Lambda\to0$. This suggests that the crossover in principle carries over to higher dimensions, but the divergent terms may have different behavior. We hope to address this in later work.

\begin{acknowledgements}
SMC is supported by DST-INSPIRE  Fellowship offered by the Dept. of Science and Technology, Govt. of India. The work is supported by the MATRICS SERB grant. The authors thank the service personnel in India whose untiring work allowed the authors to complete this work during the COVID-19 pandemic. 
\end{acknowledgements}
\appendix
\section{Elliptic Integrals and Series Expansion}
\label{appa}
To provide an analytical insight into the two limits of $\zeta$, let us assume that $N$ is large enough for the summation to be replaced by an integral in \eqref{dbcdet}. For DBC, we can introduce $\theta=(2k-1)\pi/(4N+2)$, due to which the upper limit of the integral is $\pi/2-\pi/2N$:
\begin{align}
	\det{\sigma_{red}}&\approx \frac{1}{\pi^2}\int_0^{\frac{\pi}{2}-\frac{\pi}{2N}}\frac{d\theta}{\sqrt{\Lambda+4\cos^2{\theta}}}\int_{0}^{\frac{\pi}{2}-\frac{\pi}{2N}}\sqrt{\Lambda+4\cos^2{\theta}}d\theta\nonumber\\
	&=\frac{1}{\pi^2}\int_{0}^{\frac{\pi}{2}-\frac{\pi}{2N}}\frac{d\theta}{\sqrt{1-k^2\sin^2{\theta}}}\int_0^{\frac{\pi}{2}-\frac{\pi}{2N}}\sqrt{1-k^2\sin^2{\theta}}d\theta \nonumber\\
	&=\frac{1}{\pi^2}F\left[\frac{\pi}{2}\left(1-\frac{1}{N}\right),k\right]E\left[\frac{\pi}{2}\left(1-\frac{1}{N}\right),k\right],
\end{align}
which is a product of incomplete elliptic integrals of the first and second kind\cite{1971-Byrd.Friedman-HandbookEllipticIntegrals} whose modulus is $k^2=4/(\Lambda+4)$. Now, from the exact expression above, we may write down the series expansion in two different ways:
\begin{itemize}
	\item Expanding around $N\to\infty$ and then around $\Lambda\to0$: 
	\begin{equation}
		\det{\sigma_{red}}\sim \frac{1}{2\pi^2}\log{\frac{64}{\Lambda}}+\mathscr{O}(\Lambda\log\Lambda)
	\end{equation}
	\item Expanding around $\Lambda\to0$ and then around $N\to\infty$: 
	\begin{equation}
		\det{\sigma_{red}}\sim \frac{1}{\pi^2}\log{\frac{4N}{\pi}}+\mathscr{O}(N^{-2})
	\end{equation}
\end{itemize}
The above expressions match exactly with those obtained for the cases $\zeta\ll1$ and $\zeta\gg1$ respectively. For Neumann, we introduce $\theta=(2k-1)\pi/4N$ in \eqref{nbcdet} and replace the summation with integrals:
\begin{align}
	\det{\sigma_{red}}&\approx \frac{1}{4N^2}\left[\frac{1}{2\sqrt{\Lambda}}+\frac{2N}{\pi}\int_{0}^{\frac{\pi}{2}-\frac{\pi}{4N}}\frac{d\theta}{\sqrt{\Lambda+4\cos^2{\theta}}}\right]\left[\frac{\sqrt{\Lambda}}{2}+\frac{2N}{\pi}\int_{0}^{\frac{\pi}{2}-\frac{\pi}{4N}}\sqrt{\Lambda+4\cos^2{\theta}}d\theta\right] \nonumber\\
	&= \frac{1}{\pi^2}\left[\frac{\pi}{4N}\sqrt{1+\frac{4}{\Lambda}}+\int_{0}^{\frac{\pi}{2}-\frac{\pi}{4N}}\frac{d\theta}{\sqrt{1-k^2\sin^2{\theta}}}\right]\left[\frac{\pi}{4N}\sqrt{\frac{\Lambda}{\Lambda+4}}+\int_{0}^{\frac{\pi}{2}-\frac{\pi}{4N}}\sqrt{1-k^2\sin^2{\theta}}d\theta\right] \nonumber\\
	&= \frac{1}{\pi^2}\left[\frac{\pi}{4N}\sqrt{1+\frac{4}{\Lambda}}+F\left[\frac{\pi}{2}\left(1-\frac{1}{2N}\right),k\right]\right]\left[\frac{\pi}{4N}\sqrt{\frac{\Lambda}{\Lambda+4}}+E\left[\frac{\pi}{2}\left(1-\frac{1}{2N}\right),k\right]\right],
\end{align}
where $k^2=4/(\Lambda+4)$ is the modulus of the incomplete elliptic integrals $F$ and $E$. The series expansion can be written down in two different ways:
\begin{itemize}
	\item Expanding around $N\to\infty$ and then around $\Lambda\to0$: 
	\begin{equation}
		\det{\sigma_{red}}\sim \frac{1}{2\pi^2}\log{\frac{64}{\Lambda}}+\mathscr{O}(\Lambda\log\Lambda)
	\end{equation}
	\item Expanding around $\Lambda\to0$ and then around $N\to\infty$: 
	\begin{equation}
		\det{\sigma_{red}}\sim \frac{1}{2\pi N\sqrt{\Lambda}}+\frac{1}{\pi^2}\log{\frac{8N}{\pi}}+\mathscr{O}(N^{-2})
	\end{equation}
\end{itemize}
The above expressions match exactly with the cases $\zeta\ll1$ and $\zeta\gg1$ respectively. Similarly, for periodic boundary conditions, we have:
\begin{multline}
	\det{\sigma_{red}}=\frac{1}{\pi^2}\left[\frac{\pi}{4N}\left(1+\sqrt{1+\frac{4}{\Lambda}}\right)-F\left[\frac{\pi}{2}\left(1-\frac{1}{N}\right),k\right]\right]\\\cross\left[\frac{\pi}{4N}\left(1+\sqrt{\frac{\Lambda}{\Lambda+4}}\right)-E\left[\frac{\pi}{2}\left(1-\frac{1}{N}\right),k\right]\right],
\end{multline}
where $k^2=4/(\Lambda+4)$ is the modulus of the incomplete elliptic integrals $F$ and $E$. The series expansion can be written down in two different ways:
\begin{itemize}
	\item Expanding around $N\to\infty$ and then around $\Lambda\to0$: 
	\begin{equation}
		\det{\sigma_{red}}\sim \frac{1}{2\pi^2}\log{\frac{64}{\Lambda}}+\mathscr{O}(\Lambda\log\Lambda)
	\end{equation}
	\item Expanding around $\Lambda\to0$ and then around $N\to\infty$: 
	\begin{equation}
		\det{\sigma_{red}}\sim \frac{1}{2\pi N\sqrt{\Lambda}}+\frac{1}{\pi^2}\log{\frac{4N}{\pi}}+\mathscr{O}(N^{-2})
	\end{equation}
\end{itemize}
The above expressions match exactly with the cases $\zeta\ll1$ and $\zeta\gg1$ respectively.

\section{Relation between $S$ and $\mathcal{E_N}$ in the large $N$ limit}\label{appEN}

\subsection*{Eigenvalues for logarithmic negativity}
The covariance matrix for the pure state in the case of N coupled harmonic oscillator system is given as \cite{2002Audenaert.etalPhysicalReviewA}
\begin{equation}
	\gamma=\frac{1}{2}
	\begin{bmatrix} 
		V^{-\frac{1}{2}} & 0 \\
		0 & V^{\frac{1}{2}} \\
	\end{bmatrix}
\end{equation}
where $V$ is the potential matrix for the system and 
$V^{-\frac{1}{2}}=\gamma_x$ and $V^{\frac{1}{2}}=\gamma_p$. After partial transpose we will get the covariance matrix as $\gamma^{\Gamma}$ which is defined as 
\cite{2002Audenaert.etalPhysicalReviewA} 
\begin{equation}
	\gamma^{\Gamma} = \frac{1}{2}P\gamma P
\end{equation}
where $P$ is 
\begin{equation}
	P = 
	\begin{bmatrix} 
		1 & 0 \\
		0 & -1 \\
	\end{bmatrix} 
\end{equation}
Upon using the above eq. of $\gamma^{\Gamma}$ and $P$, we finally get 
\begin{equation}
	\gamma^{\Gamma} = \frac{1}{2}
	\begin{bmatrix} 
		\gamma_x & 0 \\
		0 & \gamma_p \\
	\end{bmatrix} 
\end{equation}
which implies 
\begin{equation}
	\gamma^{\Gamma} = \gamma 
\end{equation}
Since the covariance matrix remains unchanged after the partial transpose, so 
$\mathrm{Det}(\gamma) = \mathrm{Det}(\gamma^{T})$ and hence we can use the same eigenvalues for both the entanglement entropy and logarithmic negativity for pure states.
\subsection*{Expression for logarithmic negativity}
Entanglement entropy for the finite $N$ case is given as
\cite{2014-Mallayya.etal-Phys.Rev.D}
\begin{equation}
	S = \sum_{k=1}^{m}\left(\alpha_k + \frac{1}{2}\right) \log\left(\alpha_k + \frac{1}{2}\right) - \left(\alpha_k - \frac{1}{2}\right)\log\left(\alpha_k - \frac{1}{2}\right)  
\end{equation}
where trace is taken over $m<N$ oscillators.

In the large $N$ limit $\alpha_K\rightarrow\infty$ because of zero-modes \cite{2020Chandran.ShankaranarayananPhys.Rev.D}, so we can then approximate the entanglement entropy as $S\sim\sum_k\log(\alpha_k)$ where $\alpha_k$ is the eigenvalue coming from the covariance matrix. Since we are considering only the $N$th oscillator, we simply need to consider $\alpha_N$. So, for our purpose, $S\sim\log\alpha_N$.

In the case of logarithmic negativity, we have \cite{2002Audenaert.etalPhysicalReviewA}
\begin{equation}
	\mathcal{E_N}= -\sum_{j=1}^{n}\log_2(\mathrm{min}(1,\lambda_j(Q))) 
\end{equation}
Now, the above eq. says that the $\mathcal{E_N}$ is the sum of the -ve eigenvalues of $Q$ so therefore we 
can re-write it as the sum of the absolute values of the eigenvalues of $Q$
\begin{equation}
	\mathcal{E_N}=\sum_j\log|\lambda_j|
\end{equation}
Since we are considering only the $N$th oscillator and the eigenvalues remains the same for both the entropy and negativity so we can finally express the above eq. as 
\begin{equation}
	\mathcal{E_N}=\log\alpha_N
\end{equation}
which finally leads to the fact that for pure states
\begin{equation}
	S=\mathcal{E_N}=\log\alpha_N
\end{equation}

%

\end{document}